I. **Title:**

Bond formation at polycarbonate | X interfaces (X = $Al_2O_3$, $TiO_2$, $TiAlO_2$) studied by theory and experiments

II. **Authors:**

Lena Patterer[1]\*, Pavel Ondračka[2], Dimitri Bogdanovski[1], Stanislav Mráz[1],

Peter J. Pöllmann[1], Soheil Karimi Aghda[1], Petr Vašina[2], Jochen M. Schneider[1]

[1] *Materials Chemistry, RWTH Aachen University, Kopernikusstr. 10, 52074 Aachen, Germany*

[2] *Department of Physical Electronics, Faculty of Science, Masaryk University, Kotlářská 2, 611 37 Brno, Czech Republic*

\*Corresponding author: patterer@mch.rwth-aachen.de

III. **Keywords (3-5):** polycarbonate, metal oxides, sputter deposition, *ab initio* molecular dynamics, X-ray photoelectron spectroscopy, density functional theory

IV. **Highlights:**

- The bond density at PC | X interfaces (X = $Al_2O_3$, $TiO_2$, $TiAlO_2$) was probed by XPS.
- The bond strength of interfacial bonds was determined by DFT calculations.
- O forms the most numerous and strongest interfacial bonds during sputter deposition.
- $Al_2O_3$ forms the strongest interface with PC compared to $TiO_2$ and $TiAlO_2$.

**Abstract**

Interfacial bond formation during sputter deposition of metal oxide thin films onto polycarbonate (PC) is investigated by *ab initio* molecular dynamics simulations and X-ray photoelectron spectroscopy (XPS) analysis of PC | X interfaces (X = $Al_2O_3$, $TiO_2$, $TiAlO_2$). Generally, the predicted bond formation is consistent with the experimental data. For all three interfaces, the majority of bonds identified by XPS are (C-O)-metal bonds, whereas C-metal bonds are the minority. Compared to the PC | $Al_2O_3$ interface, the PC | $TiO_2$ and PC | $TiAlO_2$ interfaces exhibit a reduction in the measured interfacial bond density by ~ 75 and ~ 65%, respectively. Multiplying the predicted bond strength with the corresponding experimentally determined interfacial bond density shows that $Al_2O_3$ exhibits the strongest interface with PC, while $TiO_2$ and $TiAlO_2$ exhibit ~ 70 and ~ 60% weaker interfaces, respectively. This can be understood by considering the complex interplay between the metal oxide composition, the bond strength as well as the population of bonds that are formed across the interface.

*Graphical abstract*

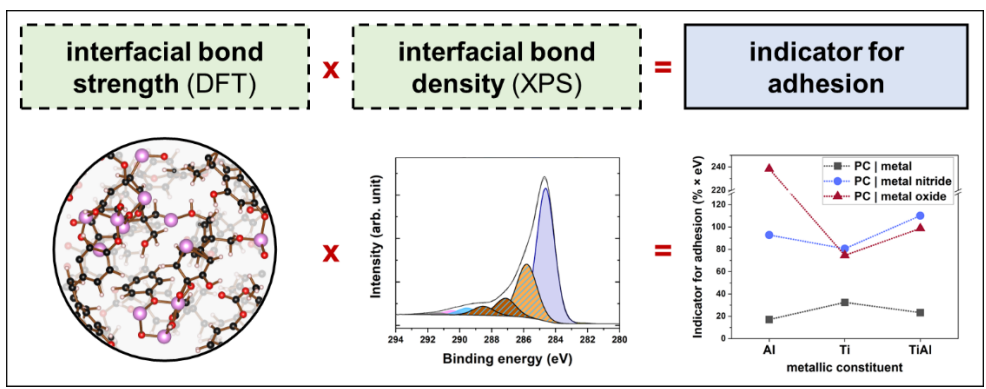

## 1. Introduction

Poly(bisphenol A carbonate), generally referred to as polycarbonate (PC), is a thermoplastic polymer with many applications, for example, in the automotive industry, the medical sector, or in electrical components [1]. However, since PC is prone to stress corrosion cracking and sensitive to scratches, surface protection by thin films is often required [2].

Amorphous alumina ($Al_2O_3$) and titania ($TiO_2$) are suitable protective thin films due to their corrosion resistance [3,4] and significantly better mechanical properties ($E$ ~ 80–180 GPa) [5,6] compared to the rather soft PC ($E$ ~ 2 GPa) [7]. Besides, both amorphous $Al_2O_3$ and $TiO_2$ have the advantage of being transparent, which widens their range of application as protective thin films for PC when transparency is required, such as safety goggles [1] or bulletproof windows [8]. Amorphous $Al_2O_3$ is already commonly used as a thin film in optical applications, such as optical lenses, antireflective coatings, and windows [9], while amorphous $TiO_2$ is considered for applications in optoelectronic devices [10] or solar cells [11] and has been reported to be highly ductile if sufficiently dense and free of geometrical flaws [12].

Protective thin films require good adhesion to the substrate to be effective and the chemical bonding at the interface between PC substrates and thin films plays a crucial role in quantifying the adhesion properties [13]. Generally, PC consists of different functional groups which are potential interaction sites during the deposition of the thin films: the (hydro-)carbon aromatic and aliphatic groups (C-C, C=C, C-H) as well as the carbonate group (O-(C=O)-O) [14].

A previous study demonstrated that the atomic layer deposition of amorphous $Al_2O_3$ and $TiO_2$ adhesion layers onto poly(methyl methacrylate) (PMMA) improved the adhesion of subsequently deposited Ti significantly when both layer thicknesses exceeded 33 nm [15]. Additionally, Gerenser [16] investigated the adhesion effect of plasma treatments prior to the deposition of Ag thin films onto polyethylene and presented the following adhesion trend of the used working gases: untreated < Ar < $O_2$ < $N_2$. This study showed that O and N play an important role in interfacial bond formation by creating new reactive $CO_x$ and $CN_x$ groups [16]. However, Amor *et al.* [17] compared the X-ray photoelectron spectroscopy (XPS) spectra of an untreated and a $CO_2$-plasma-treated PMMA | $Al_2O_3$ interface (both with an alumina film thickness of ~ 1 nm) and showed that both samples exhibit a similar C 1s spectrum with respect to new interfacial $CO_x$ groups. These data suggest that a strongly adhering interface between the polymer and thin film does not necessarily require a prior plasma treatment in case of a metal oxide deposition. This is also supported by an interfacial XPS study of $Al_2O_3$ and $ZnO_2$ deposited onto PC [18], demonstrating that upon reactive sputter deposition of a 1 nm thin film, new C-O and C=O groups are formed at both PC interfaces. However, more new C-O groups were introduced at the PC | $Al_2O_3$ interface, while more C=O groups were formed at the PC | $ZnO_2$ interface [18]. Therefore, the chemical composition of the metal oxide seems to be relevant for the interfacial group formation.

Due to the lack of systematic studies published on the interfacial bond formation of sputter-deposited alumina, titania, and a solid solution thereof onto PC, we present in the following an interfacial bond strength and population analysis of PC | X interfaces (X = $Al_2O_3$, $TiO_2$, $TiAlO_2$) by using *ab initio* simulations, electronic-structure-based bonding analyses and XPS experiments.

## 2. Methods

### 2.1. Experimental methods

PC substrates were prepared by using 5 wt.% PC pellets (additive-free, product # 4315139, Sigma Aldrich) dissolved in tetrahydrofuran (anhydrous, 99.9% purity, inhibitor-free, Sigma Aldrich). The solution was spin-coated onto 10 × 10 mm$^2$ fused-silica substrates (Siegert Wafer GmbH) by applying a velocity of 2000 rpm. Immediately after preparation, the substrates were mounted into the evacuated deposition chamber to reduce the influence of surface contamination. Additionally, thin films deposited onto conductive 10 × 10 mm$^2$ Si (100) substrates (Crystal GmbH) were analyzed regarding the film thickness, and thus, deposition rate, as well as the chemical composition.

The utilized Ti, Al, and composite Ti$_{50}$Al$_{50}$ deposition targets (purity ≥ 99.995% in each case) had a diameter of 50 mm and were mounted with a target-to-substrate distance of 100 mm, while the angle between target normal and substrate normal was 45°. The base pressure before the deposition was always ≤ 7 × 10$^{-5}$ Pa, and the Ar and O$_2$ partial pressures (both gases ≥ 99.99% purity) were set during the respective sputter deposition according to **Table 1** in order to obtain the aimed stoichiometric compositions. To ensure a homogenous deposition, the substrate was rotated at 28 rpm during deposition. No intentional substrate heating was applied.

**Table 1.** Ar and O$_2$ partial pressures during different metal oxide depositions

| Partial pressure | Al$_2$O$_3$ | TiO$_2$ | TiAlO$_2$ |
|---|---|---|---|
| $p_{Ar}$ (Pa) | 0.53 | 0.53 | 0.68 |
| $p_{O2}$ (Pa) | 0.08 | 0.03 | 0.03 |

While pulsed direct current magnetron sputtering (PDCMS) was used for the $Al_2O_3$ and $TiAlO_2$ deposition (frequency = 250 kHz, $t_{OFF}$ = 1616 ns) to avoid arcing at the target, the $TiO_2$ deposition was run in DC mode. The time-averaged power density applied to each target was 10.2 W cm$^{-2}$. Before each deposition, target sputter cleaning was performed behind a closed shutter for ≥ 2 min.

After deposition, the samples were transferred from the deposition chamber to the XPS load-lock chamber with an atmosphere exposure time of < 5 min. XPS measurements were carried out using an AXIS Supra (Kratos Analytical Ltd.) equipped with a monochromatic Al-Kα source and a hemispherical detector. The base pressure during acquisition was always < 2 × 10$^{-6}$ Pa. High-resolution spectra (C 1s, O 1s) were acquired using a pass energy of 20 eV (step size 0.05 eV, dwell time 100 ms, 20 sweeps), while for survey scans, a pass energy of 160 eV was employed (step size 0.25 eV, dwell time 100 ms, 5 sweeps). To avoid charging effects, charge neutralization was applied using a low-energy, electron-only source, and the binding energy (BE) scale was calibrated with respect to the hydrocarbon signal of PC at 284.6 eV [19]. The analysis of the XPS data was conducted using the CasaXPS software package 2.3.15 (Casa Software Ltd.), subtracting a Shirley background [20] and applying the manufacturer's sensitivity factors [21] for chemical quantification. The components of the C 1s and O 1s spectrum were fitted with a Gaussian-Lorentzian (70%-30%) line shape and the full width at half maximum (FWHM) of the C 1s components was constrained to ≤ 1.5 eV.

The analysis of the C 1s spectra aims to quantify the interfacial components and in order to compare the interfacial bond density among the different thin film systems, interface samples after the deposition of ~ 1 nm film thickness onto PC are considered. The analysis of the O 1s spectra, however, aims to elucidate the progress of the thin film growth relative to the interface formation, and therefore, interfaces with a lower film thickness are considered (≤ 0.2 nm). These thin films were obtained by using fast-acting shutters in front of the targets with an actuation time of 200 ms.

Thickness measurements were performed using an FEI Helios Nanolab 660 equipped with a field emission microscope to analyze the cross-sections of thin films deposited for 30 min onto Si. For the SEM image acquisition, an acceleration voltage and current of 10 kV and 50 pA were applied, respectively.

To determine the chemical composition of the thin films deposited for 30 min onto Si, energy dispersive X-ray spectroscopy (EDX) was conducted by using a TM4000Plus Tabletop scanning electron microscope (SEM) (Hitachi Ltd.) equipped with a Quantax75 detector and applying a 10 kV acceleration voltage. For the quantification of the $Al_2O_3$ thin film, an $Al_2O_3$ (0001) substrate (Crystal GmbH) was used as a standard, whereas the $TiO_2$ and $TiAlO_2$ thin films were quantified standardless.

### 2.2. Computational methods

Simulations of $Al_2O_3$ and $TiO_2$ sputter depositions onto PC were conducted by density functional theory (DFT)-based *ab initio* molecular dynamics (AIMD) using the OpenMX package [22–25] using the PBE exchange-correlation potential [26] and employing DFT-D3 dispersion correction [27] for an accurate representation of the van-der-Waals interactions. Methodological details on the creation of the PC bulk model consisting of 396 atoms are given in [28]. For the sputter-deposition simulation, the PC surface was bombarded by 30 atoms in total with each having a kinetic energy of 1 eV (kinetic energy of DCMS-sputtered atoms is typically in the range of a few eV [29]). While an atomic sequence of O-Al-O-O-Al in terms of incident atoms was used for the $Al_2O_3$ deposition simulation, a sequence of Ti-O-O was applied for $TiO_2$. The time interval between the incident atoms was 500 fs and the temperature during the simulation was kept constant at 300 K (close to room temperature). After the deposition of 30 atoms, the final configuration was relaxed and subsequently utilized to calculate the C 1s and O 1s core electron binding energies (BEs) by using the core-hole approach [30]. More details on the calculations can be found elsewhere [28] and all calculation data are available in the NOMAD archive [31,32].

Bonding analysis was carried out by crystal orbital Hamilton population (COHP) [33] calculations, using the LOBSTER package (version 4.0.0) [34–36] for projection of atom-resolved, local orbitals from delocalized plane-wave basis sets. The required wavefunctions which were post-processed with LOBSTER were obtained from static simulation runs using the Vienna *ab initio* simulation package (VASP, version 5.4.4) [37–39]. To prepare the required structural model, the final structure of the respective AIMD simulation (PC + deposited atoms) was excised with a distance of 13 Å from the basal plane to reduce the system size and to ensure acceptable computational time and costs. Using this simplified structure, the static VASP simulation run provided the wavefunction from which COHP calculations via LOBSTER obtained atom-resolved orbital information, and thus, the bonding characteristics. The integrated COHP values (ICOHP) per bond were extracted and interpreted as an indirect, but strongly correlated indicator of bond strength [40,41]. More computational details on the ICOHP calculation methodology can be found in [28].

## 3. Results and discussion

### 3.1. Thickness calibration and chemical composition of thin films

The thin films deposited for 30 min onto Si substrates were analyzed with respect to their thickness and chemical composition. Measuring the film thickness of SEM cross-sections, the nominal deposition rates of ~ 2.8, ~ 16.7, and ~ 20.7 nm min$^{-1}$ were determined for $Al_2O_3$, $TiO_2$, and $TiAlO_2$, respectively. To obtain thin films with a thickness of ~ 1 nm, a deposition time of 23 s, 4 s, and 3 s was utilized for $Al_2O_3$, $TiO_2$, and $TiAlO_2$ onto PC, respectively.

The chemical composition of the thin films was determined by EDX and is shown in **Table 2**. While the difference relative to the expected stoichiometry is only ~ 4% for the standard-based $Al_2O_3$ measurement, the standardless measurements of $TiO_2$ and $TiAlO_2$ exhibit a greater deviation with ≤ 13%, which is, however, still within the systematic uncertainty of EDX [42].

**Table 2.** Chemical composition of thin films deposited for 30 min onto Si (EDX).

|           | Al (at.%) | Ti (at.%) | O (at.%) |
|-----------|-----------|-----------|----------|
| $Al_2O_3$ | 39        | --        | 61       |
| $TiO_2$   | --        | 37        | 63       |
| $TiAlO_2$ | 27        | 26        | 47       |

### 3.2. Simulated sputter deposition of $Al_2O_3$ and $TiO_2$ onto PC

For the *ab initio* simulations, only the interfaces of PC with $Al_2O_3$ and $TiO_2$ are considered and compared to the pristine PC surface (**Figure 1a**). **Figure 1b** & **c** shows the plan view of the interfaces after the simulated sputter deposition of 6 formula units of $Al_2O_3$ and 10 formula units of $TiO_2$ onto PC, respectively (30 atoms in total for each case).

Comparing the interfaces with the pristine PC surface (**Figure 1a**), a significantly higher fraction of C-O groups is observed, indicating reactions of incident O atoms with the polymer. Additionally, C-metal bonds are visible, corresponding to the reaction of C atoms with both metal species, Al and Ti (**Figure 1b** & **c**). Due to the different stoichiometry of $Al_2O_3$ and $TiO_2$, the oxygen-to-metal ratio is naturally larger at the PC | $TiO_2$ interface.

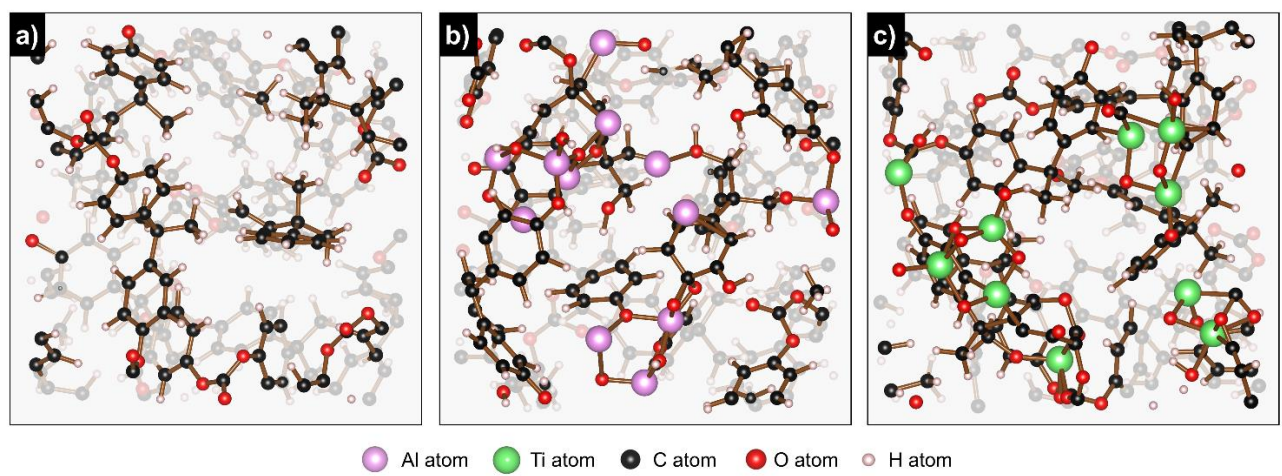

**Figure 1.** Plan view of a) pristine PC [28], b) the PC | $Al_2O_3$ interface, and c) the PC | $TiO_2$ interface.

*3.3.    Comparison of calculated and experimental C 1s spectra*

When analyzing the XPS survey scan of pristine PC, a surface concentration of ~ 84 at.% C and ~ 16 at.% O is measured which is in good agreement with values from literature (C/O = 83/17) [19]. The calculated and measured C 1s spectra of pristine PC are depicted in **Figure 2a** & **b**, respectively, and both show signals of the C=C, C-C, C-H (blue), the $C_{ring}$-O (orange), and the O-(C=O)-O (pink) groups. The BEs of the groups detected by XPS are in good agreement with BE values from literature [14,19], including the two π-π* shake-up satellite signals detected at 291.5 and 292.5 eV associated with aromatic bonding [19] (**Figure 2b**). Also, the calculated BEs of the (hydro-)carbon and the $C_{ring}$-O groups at 284.6 eV and 286.2 eV, respectively, agree well with the XPS C 1s spectrum, whereas the BE shift of the carbonate group (O-(C=O)-O) relative to the hydrocarbon group is underestimated by DFT ($\Delta BE_{DFT}$ = 4.7 eV vs. $\Delta BE_{XPS}$ = 5.8 eV) (**Figure 2a** & **b**). While differences of 1 eV are to be expected for BEs calculated by DFT [43,44], the trends of the chemical shifts are consistent with those in the experimental spectrum.

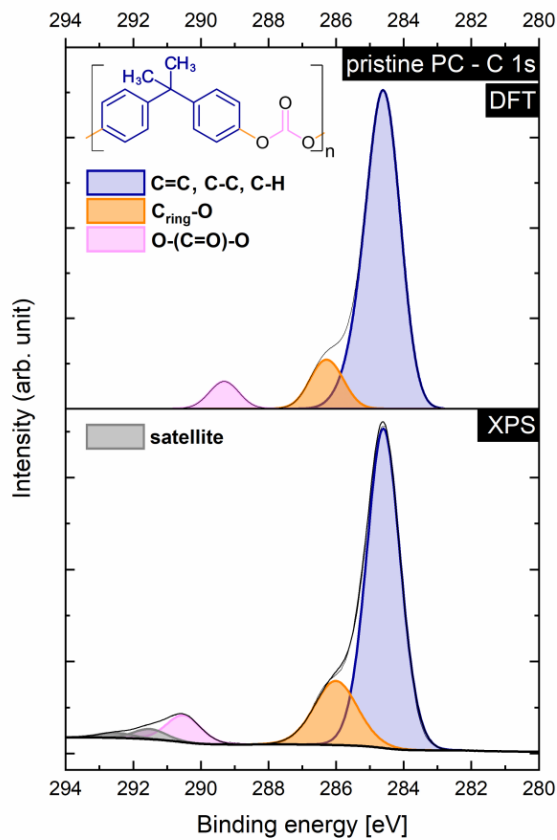

**Figure 2.** C 1s spectra of pristine PC a) calculated by DFT (calculated from the PC bulk model depicted in **Figure 1**a), and b) measured by XPS (both spectra are adapted from our previous study [28]).

As C is, besides H, the major constituent of PC, the changes in the C 1s spectrum play an important role in probing the interfacial bonds forming with sputter-deposited thin-film atoms. By comparing the C 1s spectra of the PC | $Al_2O_3$ and PC | $TiO_2$ interfaces to

that of pristine PC, chemical state changes of C atoms upon deposition are investigated. The theoretical and experimental C 1s spectra of the PC | $Al_2O_3$ interface are depicted in **Figure 3a** and **b,** respectively. The simulations predict interface formation via both interfacial (C-O)-Al groups (light blue, **Figure 3a**) and C-Al (yellow, **Figure 3a**). Experimentally (**Figure 3b**), the formation of (C-O)-Al and $CO_x$ groups is verified by three components detected between 285.8-289.5 eV (light blue, light-blue/brown striped), while the C-Al component is detected at 282.2 eV (yellow). To analyze the effect of pure O on the interfacial bond formation, an O-plasma treatment of PC substrates was carried out and the resulting C 1s spectrum is shown in the supporting information (**Figure S1**). The C 1s spectrum of the O-plasma-treated PC looks very similar to the one coated by $Al_2O_3$ confirming that most interfacial bonds are formed by (C-O)-Al and $CO_x$ groups, connecting PC and the $Al_2O_3$ thin film. Generally, a higher BE of $CO_x$ groups means more O bonds attached to the C atom (e.g. BE (C=O) ~ 287.9 eV, BE (O-C=O) ~ 289.0 eV [17]). However, one difference between the C 1s spectra of the O-plasma treated PC and PC | $Al_2O_3$ is the blue (C-O)-Al component at BE = 289.6 eV, indicating the reaction of the pristine carbonate group (O-(C=O)-O) with Al atoms as observed previously [45].

Considering the experimentally determined population of interfacial (C-O)-Al + $CO_x$ groups (**Figure 3b**), indicated by the area fraction of the light blue and light-blue/brown components, it is ~ 80 times larger than the area fraction of the C-Al groups (yellow component). This is in contrast to the calculated C 1s spectrum, which predicts nearly the same population of C-Al and (C-O)-Al groups (**Figure 3a**) and may indicate that the O concentration at the interface is significantly higher compared to the stoichiometric $Al_2O_3$ ratio assumed for the simulations. Analyzing the survey scan after a 4 s-deposition of $Al_2O_3$ onto PC (thickness ~ 0.2 nm) indicates a concentration of deposited O at the interface that is ~ 4.5 times higher than that of Al (the stoichiometric ratio assumed for the simulation is O/Al = 1.5). A similar experimental observation was made for the metal nitride depositions showing a higher fraction of C-N compared to C-metal groups [28]. Hence, while these simulations predict which interfacial groups are formed very well, the population of these groups might differ compared to the experiment due to diverging O/metal ratios at the simulated vs. experimental interfaces.

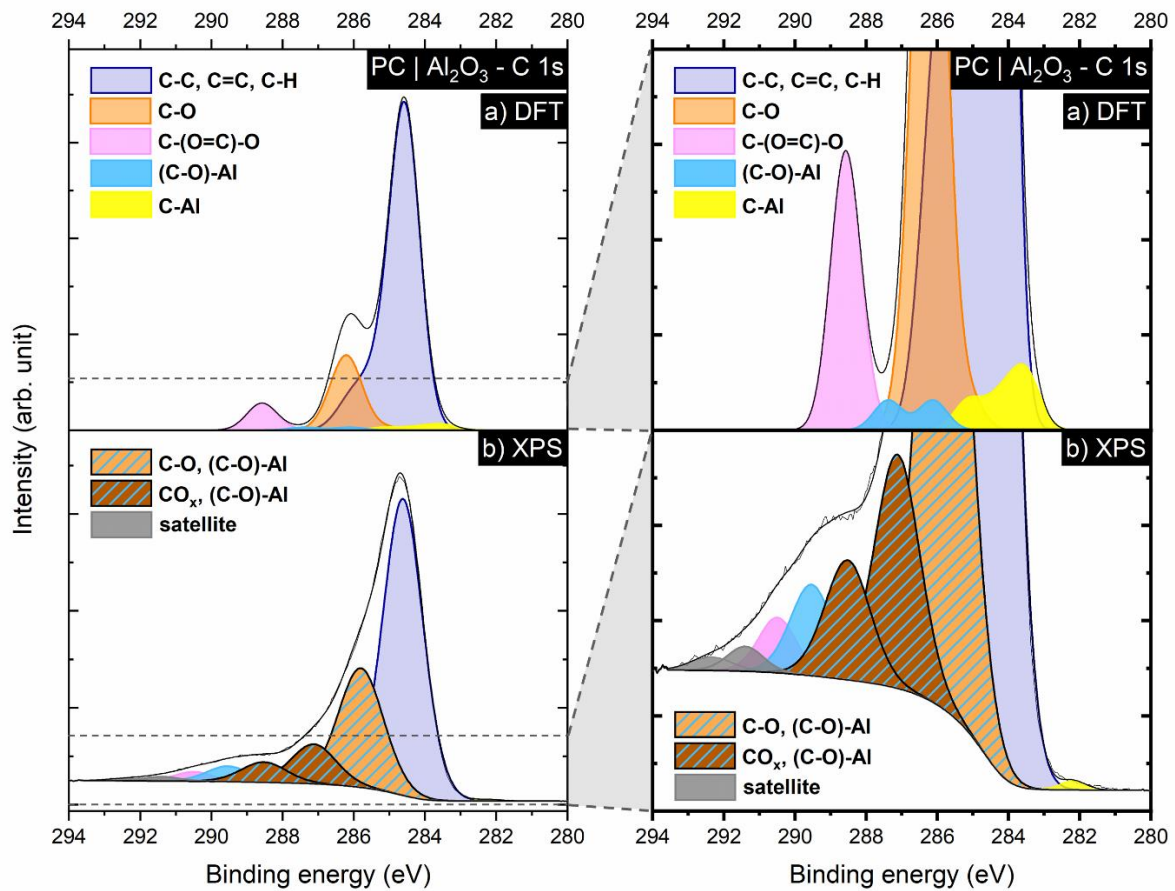

**Figure 3.** C 1s spectra a) calculated by DFT and b) measured by XPS for the PC | $Al_2O_3$ interface shown as (left) whole spectra and (right) magnification of the low-intensity C 1s components. The different groups are indicated by the color code given in the legend.

Considering the PC | TiO$_2$ interface, the calculated and experimentally detected C 1s spectra are depicted in **Figure 4a** & **b**, respectively. The PC | TiO$_2$ simulation predicts an interface formation via C-Ti bonds (green component) and (C-O)-Ti bonds (red component), while both components exhibit a similar BE of ~ 282.4.-283.9 eV and ~284.2 eV respectively (**Figure 4a**).

The experimental C 1s spectrum of the PC | TiO$_2$ interface also shows evidence of the C-Ti component at BE = 282.2 eV (**Figure 4b**); however, the numerous experimentally detected (C-O)-Ti groups with distinct signals (red, brown/red-striped, orange/red-striped) exhibit higher BEs (285.8 eV-289.1 eV) compared to the predicted groups (**Figure 4a**).

Despite the similarities of the experimental C 1s spectra of PC | Al$_2$O$_3$ and PC | TiO$_2$, it is evident that significantly more (C-O)-metal + CO$_x$ groups are formed upon Al$_2$O$_3$ deposition compared to the TiO$_2$ case. Analyzing the survey scan after the deposition of TiO$_2$ onto PC for 0.5 s (thickness ~ 0.1 nm), an O/Ti ratio of 2.5 is obtained, which is close to the stoichiometric ratio of 2 assumed for the simulations. This indicates that the initial O/metal ratio at the PC | TiO$_2$ interface (2.5) is lower compared to the one at the PC | Al$_2$O$_3$ interface (4.5).

To obtain more insights into the mechanisms of the interfacial bond formation of PC | Al$_2$O$_3$ and PC | TiO$_2$ with regard to similarities and differences, the O 1s spectra of both interfaces with thinner film thickness (< 0.2 nm) are analyzed in the next step of this study.

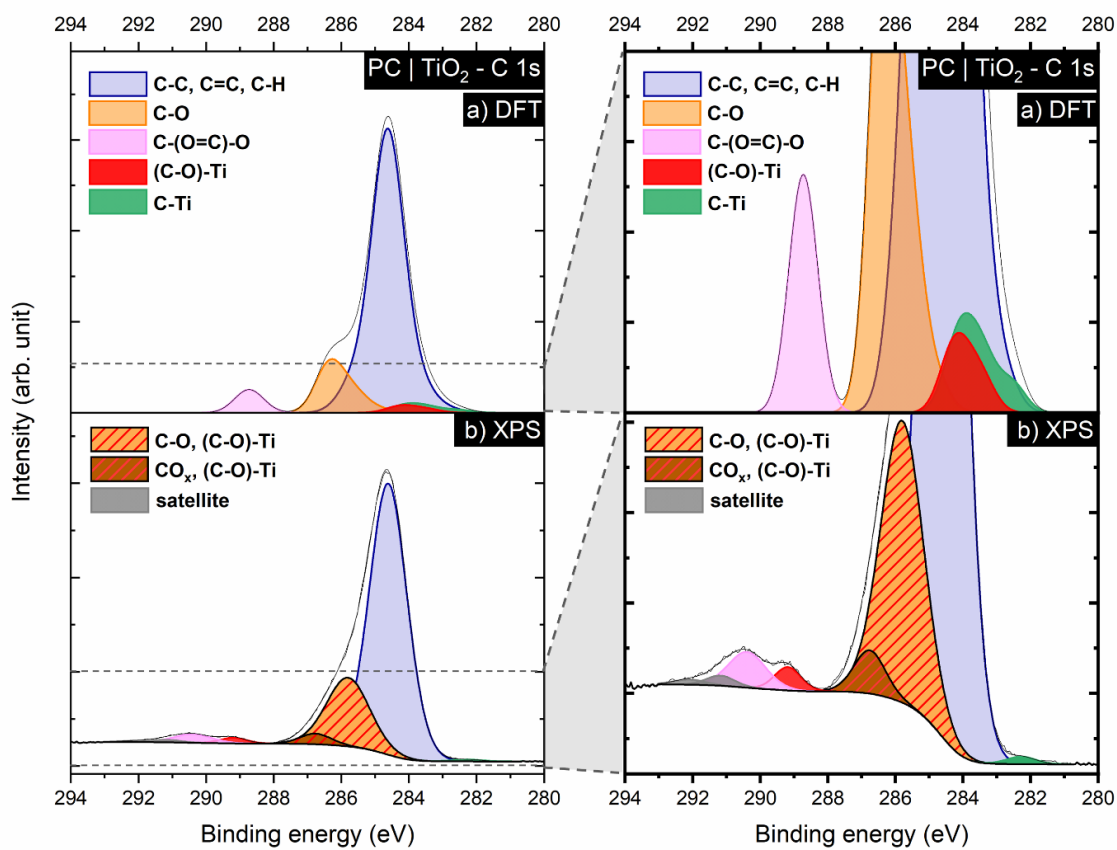

**Figure 4.** C 1s spectra a) calculated by DFT and b) measured by XPS for the PC | TiO$_2$ interface shown as (left) whole spectra and (right) magnification of the low-intensity C 1s components. The different groups are indicated by the color code given in the legend.

*3.4.    Comparison of calculated and experimental O 1s spectra*

Considering the O 1s spectra of the PC | Al$_2$O$_3$ and the PC | TiO$_2$ interfaces, more information about the interfacial O-based groups and about the growing metal oxide thin film covering the interface can be extracted. A pristine PC monomer exhibits two O 1s contributions (**Figure 5a**): one double-bonded O atom (C=O) at BE = 532.3 eV and the two single-bonded O atoms (C$_{ring}$-O-C) at BE = 533.9 eV [46].

When the sputter deposition of $Al_2O_3$ onto PC is simulated (**Figure 5b-i**), the calculated O 1s spectrum exhibits, besides the pristine O=C (pink) and $C_{ring}$-O-C groups (orange), additional contributions of interfacial C-O-Al groups (light blue), thin-film related O-Al groups within the growing thin film (purple), and various gaseous $CH_xO_y$ molecules (grey). The interfacial C-O-Al groups exhibit BEs in the range of ~ 530.3-533.8 eV, implying reactions with mostly single-bonded O (light blue contributions at > 532.2 eV) but also some reactions with double-bonded O (light blue contributions at < 532.2 eV). Both the C-O-C and the O=C groups are shifted towards lower BE upon the interaction with Al atoms due to a lower oxidation state.

The experimentally detected O 1s spectrum of PC | $Al_2O_3$ (**Figure 5b-ii**) constitutes a symmetric peak shape. Nevertheless, two distinct components are identified with the component at higher BE of ~ 533.2 eV attributed to C-O-C and C-O-Al contributions (orange/blue striped), while the component at lower BE of ~ 533.2 eV is assigned to O=C and O-Al contributions (pink/purple striped, **Figure 5b-ii**).

When $TiO_2$ is deposited onto PC, the simulation predicts two main C-O-Ti contributions at BE ~ 532.5 eV and ~ 530.8 eV (both red) indicating the reaction with the single- and double-bonded O groups, respectively (**Figure 5c-i**). When the chemical environment of O is defined exclusively by Ti atoms, a component at BE ~ 529.8 eV is determined (turquoise, **Figure 5c-i**). Additionally, the formation of various gaseous $CH_xO_y$ compounds is observed after the simulation (grey components, **Figure 5c-i**).

Overall, the predicted O 1s spectrum for PC | $TiO_2$ is wider (~ 4 eV between C-O-C and O-Ti, **Figure 5c-i**) compared to the O 1s spectrum of PC | $Al_2O_3$ (~ 2.5 eV between C-O-C and O-Al, **Figure 5b-i**). This can be explained by the larger electronegativity differences between Ti (1.54) and C (2.55) compared to Al (1.61) and C [47]. A similar

trend was previously observed for the N 1s spectra of AlN and TiN [28]. The wider O 1s spectrum of the experimental PC | $TiO_2$ allows for a better distinction of the individual O 1s contributions (**Figure 5c-ii**).

The component at the highest BE ~ 533.9 eV of the experimental O 1s spectrum is attributed to the pristine single-bonded O groups of PC (orange), whereas the largest component at BE ~530.4 eV is attributed to the growing $TiO_2$ thin film (turquoise, **Figure 5c-ii**). The component at BE ~ 531.8 eV (red/pink striped) is assigned mainly to interfacial C-O-Ti groups, as well as a small contribution of C=O groups, although the latter should constitute only ~ 50% of the area detected for the C-O-C component (compare **Figure 5a**).

When comparing the experimental O 1s spectra of both interfaces, it is evident that the O-Ti component (turquoise), corresponding to the growing thin film, constitutes a higher area fraction compared to the corresponding thin-film component of $Al_2O_3$ (pink/purple striped). Since both XPS O 1s spectra reflect the interface state after the deposition of thin films with a similar thickness (< 0.2 nm for both $Al_2O_3$ and $TiO_2$), $TiO_2$ seems to form a thin film covering the interfacial region earlier compared to $Al_2O_3$. Accordingly, the simulations predict a higher O-Ti population (turquoise component) relative to the O-Al population (purple component) after the deposition of 30 atoms (**Figure 5b-i** & **c-i** with same y-axis dimensions), thus also indicating a higher tendency of O to form O-Ti groups compared to O-Al groups.

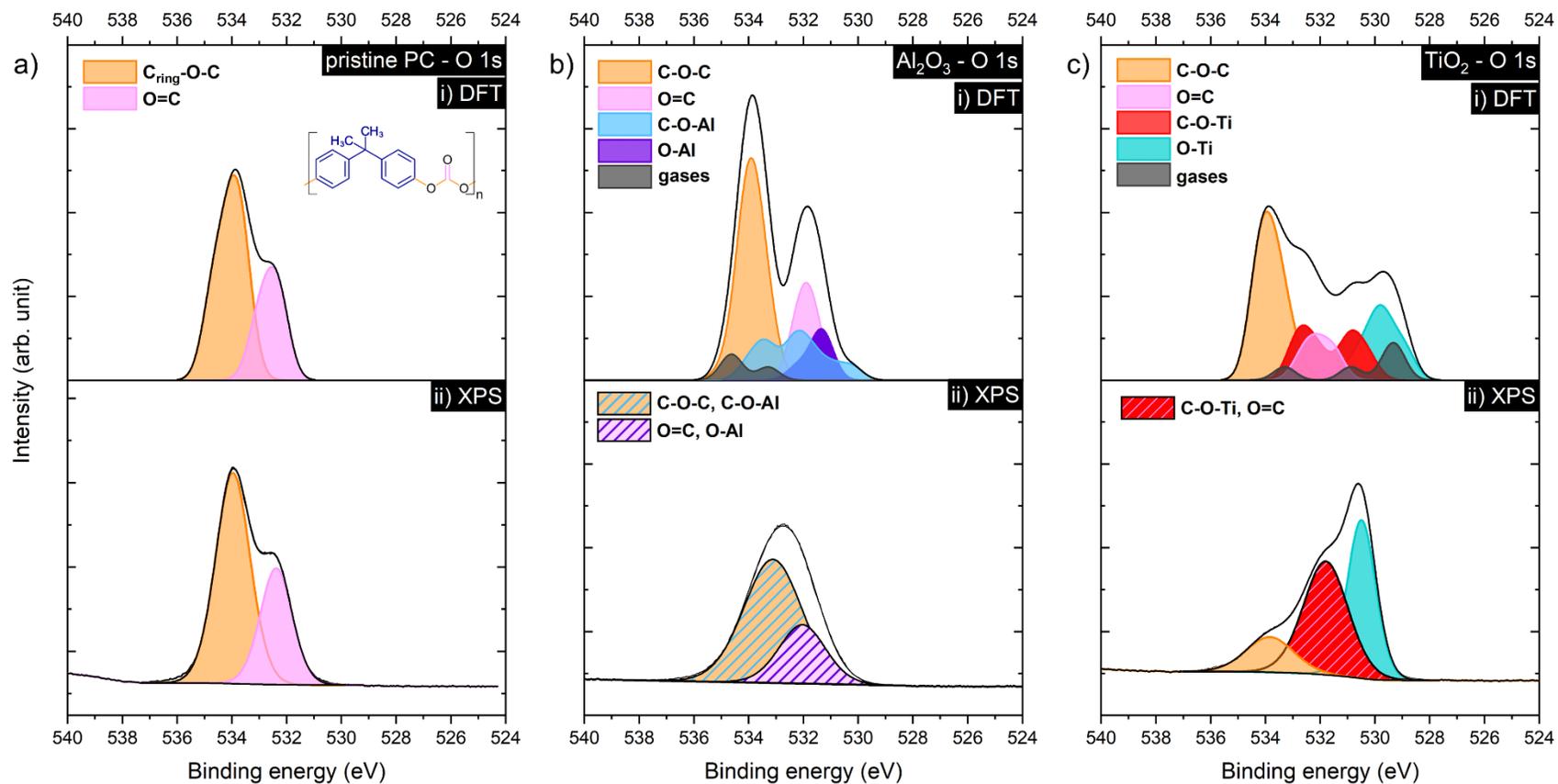

**Figure 5.** O 1s spectra of a) pristine PC, b) the PC | $Al_2O_3$ interface, and c) the PC | $TiO_2$ interface determined (i) by DFT and (ii) by XPS measurements. The different groups are indicated by the color code given in the legend.

*3.5.     Interfacial bond density*

For the experimental evaluation of the interfacial bond density, the area of the interfacial C 1s components, formed upon the deposition of a 1 nm-thick film onto PC, relative to the overall C 1s signal was determined according to our previously proposed method [28]:

$$C\ 1s\ interfacial\ component\ fraction\ (\%) = \frac{Area\ of\ interfacial\ C\ 1s\ component}{Area\ of\ whole\ C\ 1s\ signal} \times 100$$

In **Figure 6,** the *C 1s interfacial component fraction* is compared for three PC | X interfaces (X = $Al_2O_3$, $TiO_2$, $TiAlO_2$) and subcategorized into the different interfacial bond types. Here, the ternary metal oxide $TiAlO_2$ is also included in the experimental evaluation of the interfacial bond density, even though no extra simulation was conducted for this system. The corresponding XPS C 1s and O 1s spectra for the analysis of the PC | $TiAlO_2$ interface are shown in the supporting information (**Figure S2**, **Figure S3**).

**Figure 6** reveals for the PC | $Al_2O_3$ interface the highest *C 1s interfacial component fraction* or stated in a simpler term – the highest interfacial bond density –due to the formation of many (C-O)-Al bonds. In comparison, PC | $TiO_2$ and PC | $TiAlO_2$ exhibit ~ 75% and ~ 65% less interfacial bonds, respectively (**Figure 6**). Also, compared to the corresponding metallic [45] and the metal nitride systems [28], substantially more interfacial bonds are formed at the PC | $Al_2O_3$ interface (**Figure 6**).

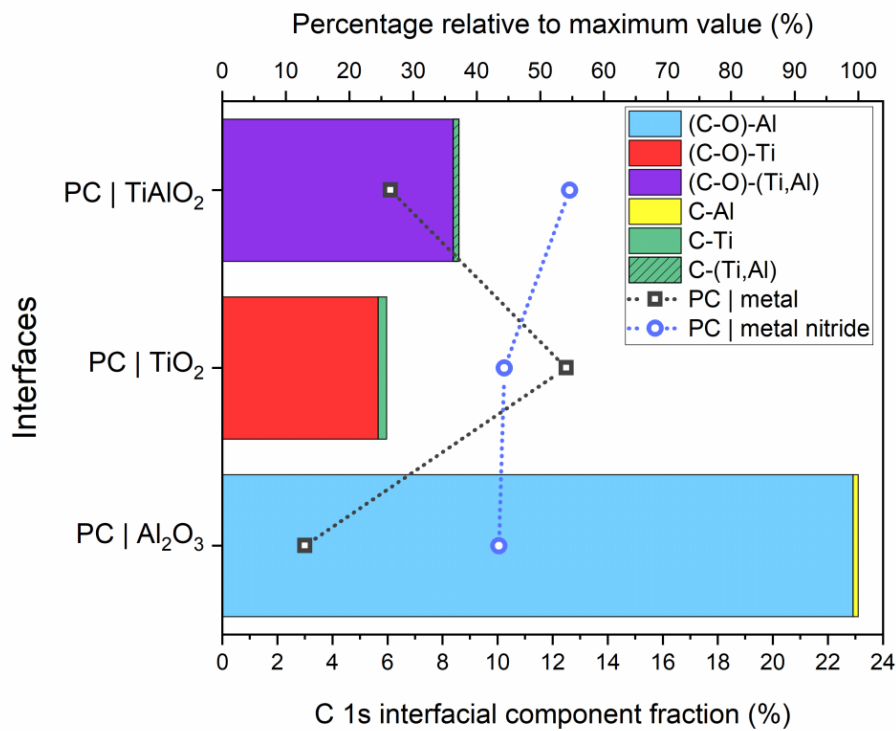

**Figure 6.** C 1s interfacial component fraction of interfacial groups determined for PC | X3 interfaces (X3 = $TiO_2$, $Al_2O_3$, $TiAlO_2$). The overall C 1s interfacial component fraction for PC | X1 (X1 = Al, Ti, TiAl) [45] and PC | X2 (X2 = AlN, TiN, TiAlN) [28] are shown as grey squares and blue circles, respectively. Lines between data points are depicted as a guide for the eye.

### 3.6. *Interfacial bond strength analysis (ICOHP)*

To evaluate the interfacial bond strength, ICOHP calculations were performed for the simulated PC | $Al_2O_3$ and PC | $TiO_2$ interfaces and the ICOHP values for the corresponding interfacial bonds are depicted as a function of bond length and interacting atoms in **Figure 7**. In general, a more negative (or larger absolute) ICOHP value implies a stronger bond. As shown in **Figure 7**, the C-metal bonds exhibit a weaker bond strength (ICOHP between −1.1 and −5.0 eV) compared to the C-(O-metal) bonds (between −8.0 and −16.1 eV). The C-(O-metal) bonds correspond to the ICOHP value between a C and an O atom, while the O atom is in this case also bonded to a metal atom, thus forming an interfacial bond between the thin film and PC.

Additionally, the simulation predicts a higher interfacial bond density for the PC | $TiO_2$ interface (22 interfacial bonds/30 deposited atoms) compared to the PC | $Al_2O_3$ interface (17 interfacial bonds/30 deposited atoms) due to the higher number of C-Ti bonds (number of green data points) compared to fewer C-Al bonds (number of yellow data points) (**Figure 7**). This deviating trend from the experimentally measured interfacial bond density (**Figure 6**) can be explained by a different initial O/metal ratio at the interface during the experimental sputter deposition as discussed before.

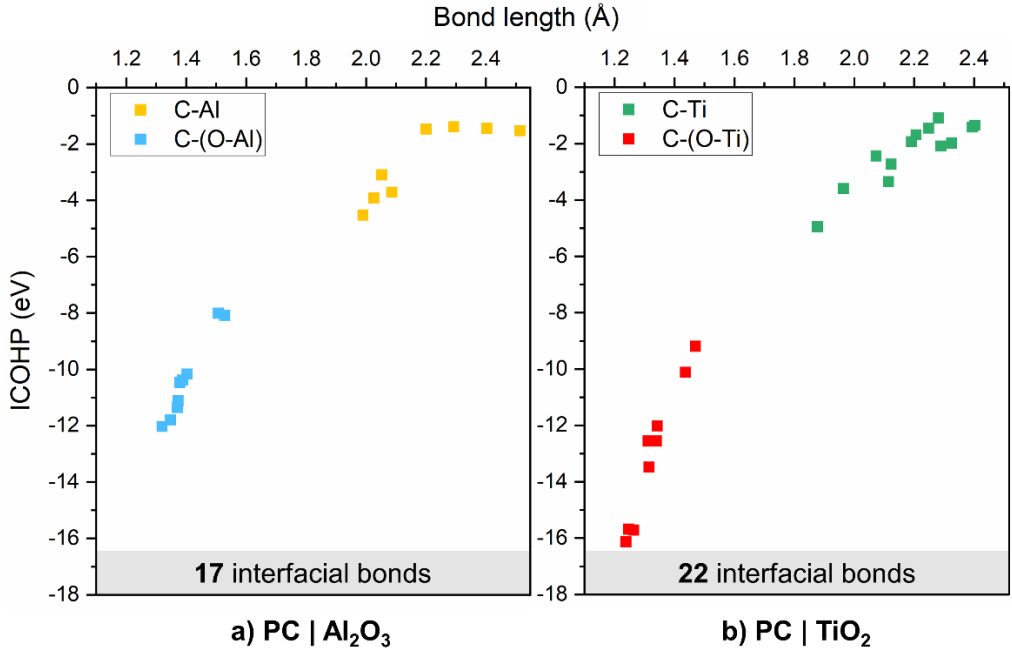

**Figure 7.** ICOHP analysis of individual interfacial bonds at the simulated a) PC | $Al_2O_3$, and b) PC | $TiO_2$ interfaces and their total number of interfacial bonds at the final state of the sputter deposition simulation.

### 3.7. Indicator for adhesion

To compare the interfacial strength, the *indicator for adhesion* was defined previously [45], corresponding to the average absolute ICOHP value for each bond type (**Figure 7**) multiplied by the experimentally determined *C 1s interfacial component fraction* (**Figure 6**). The approximate ICOHP values for the PC | $TiAlO_2$ interface were determined by averaging the interfacial C-metal and C-(O-metal) bonds at the PC | $Al_2O_3$ and the PC | $TiO_2$ interfaces. Our previous study indicated that the ICOHP values for interfacial C-N, C-metal, and (C-O)-metal bonds at the ternary PC | TiAlN interface exhibit approximately the values averaged over the dual PC | AlN and PC | TiN systems for the

respective bond type [28]. More details on the calculations of the *indicator for adhesion* can be found in the supporting information (**Table S1**).

By comparing the *indicator for adhesion* for the PC | X interfaces (X = $Al_2O_3$, $TiO_2$, $TiAlO_2$) (**Figure 8**), the strongest interfacial bond formation is determined for $Al_2O_3$ due to the numerous strong (C-O)-Al bonds (see **Figure 6**). The second strongest interface is formed for PC | $TiAlO_2$, while PC | $TiO_2$ shows the weakest interface formation (**Figure 8**). For all three interfaces, the main contributions to the *indicator for adhesion* are defined by interfacial (C-O)-metal bonds, whereas C-metal bonds play a negligible role due to their low detected density (see **Figure 6**), as well as their weaker bond strength (see **Figure 7**).

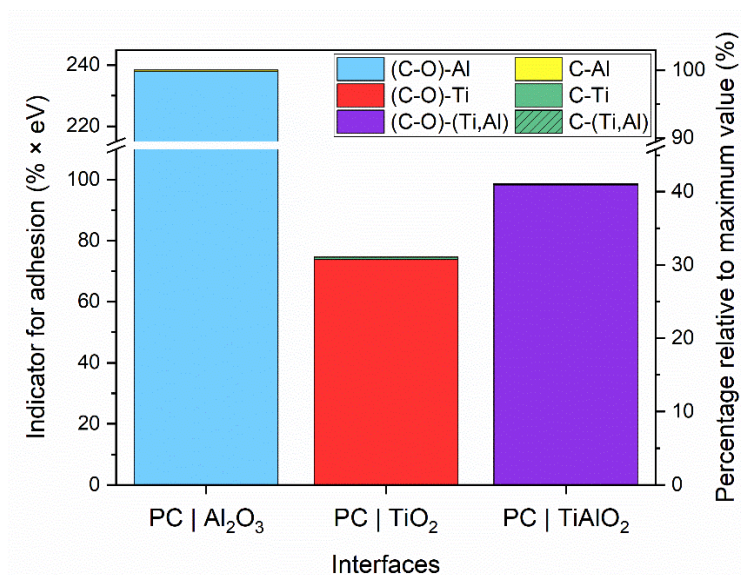

**Figure 8.** Combination of the measured relative interfacial bond density and the absolute ICOHP values of the respective bond type as an *indicator for adhesion* for the PC | X interfaces (X = $TiO_2$, $Al_2O_3$, $TiAlO_2$). An axis break between 110-215 % × eV (left axis) and 45-90% (right axis) is included for better visibility.

When comparing the overall values of the *indicator for adhesion* for the PC interface with sputter-deposited metals (Al, Ti, TiAl) [45], metal nitrides (AlN, TiN, TiAlN) [28] and metal oxides ($Al_2O_3$, $TiO_2$, $TiAlO_2$), some additional trends can be observed (**Figure 9**). Based on **Figure 9**, the metallic systems form weaker interfaces with PC compared to the metal nitride and metal oxide systems. Since PC | Ti exhibits a similar or higher interfacial bond density compared to the metal nitride and metal oxide systems (**Figure 6**), the weaker interfacial bond strength of the metal-based bonds is the reason for the lower *indicator for adhesion.* In contrast, the rather strong interfacial N-based and O-based bonds lead to stronger interfaces for the metal nitride and metal oxide systems. This observation is also in good agreement with literature reporting better adhesion for metals (Ag) deposited onto N- or O-plasma-treated polymer surfaces compared to the depositions onto untreated or Ar-treated surfaces [16].

Additionally, it is observed in **Figure 9** that TiN and $TiO_2$ as well as TiAlN and $TiAlO_2$ show similar values for the *indicator for adhesion*, while the nitride systems in both cases exhibit a slightly higher value compared to the oxide systems. This trend is in good agreement with previous reports in literature [16,48]. The simulations showed that both N and O form strong bonds at the PC interface with similar ICOHP values. However, the interfacial XPS analysis of the C 1s spectra conducted for TiN and $TiO_2$, as well as for TiAlN and $TiAlO_2$, reveal a higher concentration of N-based interfacial bonds compared to the O-based interfacial bonds. In agreement with this, the simulations indicate that a significantly higher interfacial bond density is formed after the deposition of TiN (34 bonds/30 deposited atoms) [28] compared to $TiO_2$ (22 bonds/30 deposited atoms). One explanation might be that O has - compared to N - a higher tendency to form bonds with Ti. When comparing both configurations after the simulations, only 1 out of 18 N-Ti bonds is detected that is not part of a bridging C-N-Ti group. Contrarily, 18 out of 29 O-Ti bonds are formed which are not part of a bridging C-O-Ti connection. Hence, O has a higher tendency to form bonds with Ti exclusively compared to N, and therefore, contributes less to the interfacial bond formation by forming bridging bonds with C groups of the polymer.

When comparing the trends of the *indicator for adhesion* among the metals, metal nitrides, and metal oxides, it is evident that Ti forms the strongest interface among the metals, however, for TiN and $TiO_2$, the weakest interfacial bond formation is observed among the metal nitrides and metal oxides, respectively. As already discussed in our previous study [28], it seems that the high reactivity of both Ti and N leads to fewer strong C-N bonds, while numerous weak Ti-based interfacial bonds are formed instead. The same trend seems to be valid for the $TiO_2$ systems and is enhanced by the high tendency to form O-Ti rather than bridging C-O-Ti bonds, as discussed above.

The overall highest *indicator for adhesion* was determined for PC | $Al_2O_3$, which is more than twice as high compared to the second strongest interface (PC | TiAlN). The higher *indicator for adhesion* for AlN compared to TiN was explained by the low reactivity of Al to form bonds with the hydrocarbon groups of PC, thus leading to many strong C-N bonds at the interface [28]. Similarly, the high concentration of bridging interfacial C-O is the main reason for the high *indicator for adhesion* of the PC | $Al_2O_3$.

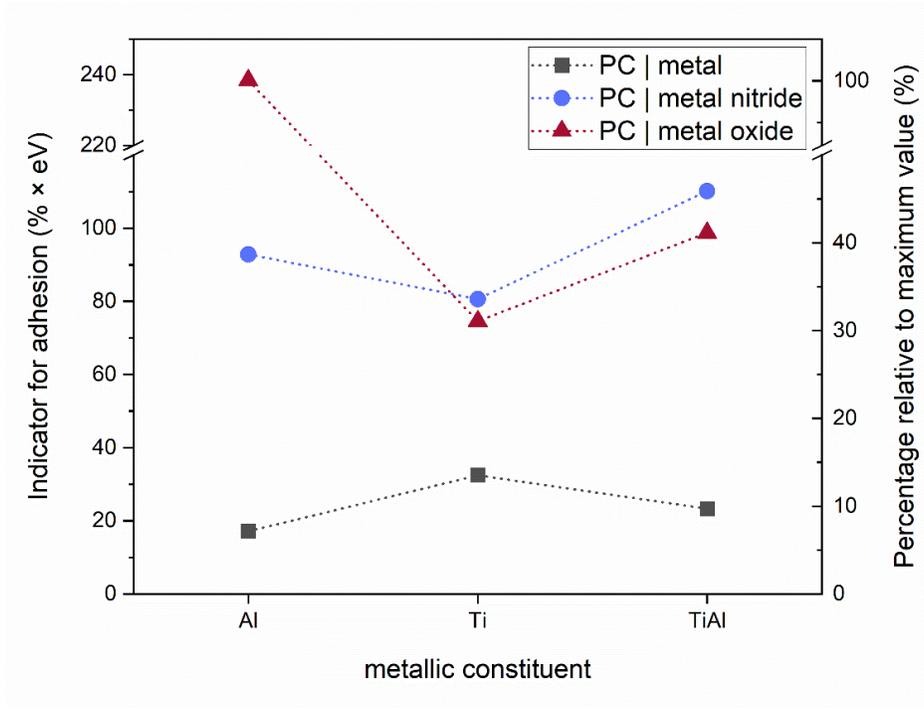

**Figure 9.** The *indicator for adhesion* for PC | X1 interfaces (X1 = Ti, Al, TiAl,) [45] is depicted as grey squares, for the PC | X2 interfaces (X2 = TiN, AlN, TiAlN) [28] as blue cycles, and for the PC | X3 interfaces (X3 = $Al_2O_3$, $TiO_2$, $TiAlO_2$) as red triangles. The *indicator for adhesion* is based on the calculated ICOHP values of the individual interfacial bond types multiplied by the interfacial bond density determined experimentally after the deposition of a film with ~ 1 nm thickness. An axis break between 120-220 % × eV (left axis) and 45-90% (right axis) is included for better visibility, while lines between symbols are depicted as a guide for the eye.

## 4. Conclusions

This study compared the interfacial bond formation of sputter-deposited $Al_2O_3$, $TiO_2$, and $TiAlO_2$ thin films onto PC substrates by correlative *ab initio* simulations and XPS measurements. While the predicted formation of interfacial C-metal and (C-O)-metal groups by simulations was confirmed by XPS, their populations deviated from predictions due to an experimentally higher initial O/metal ratio. While for the PC | $Al_2O_3$ interface a high density of (C-O)-Al groups is detected by XPS the interfacial bond density detected at the PC | $TiAlO_2$ and PC | $TiO_2$ interfaces is ~ 65 and ~ 75% lower, respectively. Due to the significantly higher concentration of interfacial bonds, also the *indicator for adhesion* - a measure combining the calculated bond strength with the XPS-measured interfacial bond density - is highest for the PC | $Al_2O_3$ interface, followed by a ~ 60% weaker PC | $TiAlO_2$ interface, as well as a ~ 70% weaker PC | $TiO_2$ interface. Comparing the *indicator of adhesion* for metal oxides ($Al_2O_3$, $TiO_2$, $TiAlO_2$), metals (Al, Ti, TiAl), and metal nitrides (AlN, TiN, TiAlN), additional mechanisms and trends are identified to form strong interfaces with PC:

- The strongest PC | metal interface is obtained for Ti, due to the largest interfacial bond density. Even though Al forms stronger interfacial bonds compared to Ti, its lower interfacial bond density leads to the weakest interface.
- Among the metal nitride systems, TiN exhibits approximately the same interfacial bond density compared to AlN, but fewer strong C-N bonds are formed leading to an overall weaker interface for PC | TiN compared to PC | AlN.
- A high initial O/metal or N/metal ratio is preferential for a strong interface formation of metal oxides or metal nitrides deposited onto PC due to the high interfacial bond density of strong C-N and C-O groups.

## 5. Acknowledgments

This research was funded by the German Research Foundation (DFG, SFB-TR 87/3) "Pulsed high power plasmas for the synthesis of nanostructured functional layers" and project LM2023039 funded by the Ministry of Education, Youth and Sports of the Czech Republic. The authors gratefully acknowledge the computing time and support provided by the IT Center of RWTH Aachen University and granted by the Jülich-Aachen Research Alliance (JARA) within the framework of the JARA0151 and JARA0221 projects.

## 6. CRediT author statement

**Lena Patterer:** Conceptualization, Methodology, Formal analysis, Investigation, Writing – original draft, Visualization. **Pavel Ondračka:** Methodology, Formal analysis, Investigation, Writing – original draft. **Dimitri Bogdanovski:** Methodology, Formal analysis, Investigation, Writing – original draft. **Stanislav Mráz:** Methodology, Investigation, Writing – review & editing. **Peter J. Pöllmann:** Formal analysis, Investigation, Writing – review & editing **Soheil Karimi Aghda:** Formal analysis, Investigation, Writing – review & editing. **Petr Vašina:** Supervision, Funding acquisition, Writing – review & editing. **Jochen M. Schneider:** Conceptualization, Methodology, Writing – original draft, Supervision, Project administration, Funding acquisition.

## 7. Supporting information

Bond formation at polycarbonate | X interfaces (X = Al$_2$O$_3$, TiO$_2$, TiAlO$_2$)
studied by theory and experiments

*Lena Patterer[1]\*, Pavel Ondračka[2], Dimitri Bogdanovski[1], Stanislav Mráz[1],
Peter J. Pöllmann[1], Soheil Karimi Aghda[1], Petr Vašina[2], Jochen M. Schneider[1]*

[1] Materials Chemistry, RWTH Aachen University, Kopernikusstr. 10, 52074 Aachen, Germany

[2] Department of Physical Electronics, Faculty of Science, Masaryk University, Kotlářská 2, 611 37 Brno, Czech Republic

\*Corresponding author: patterer@mch.rwth-aachen.de

### 7.1. XPS C 1s spectra of oxygen-plasma treated PC

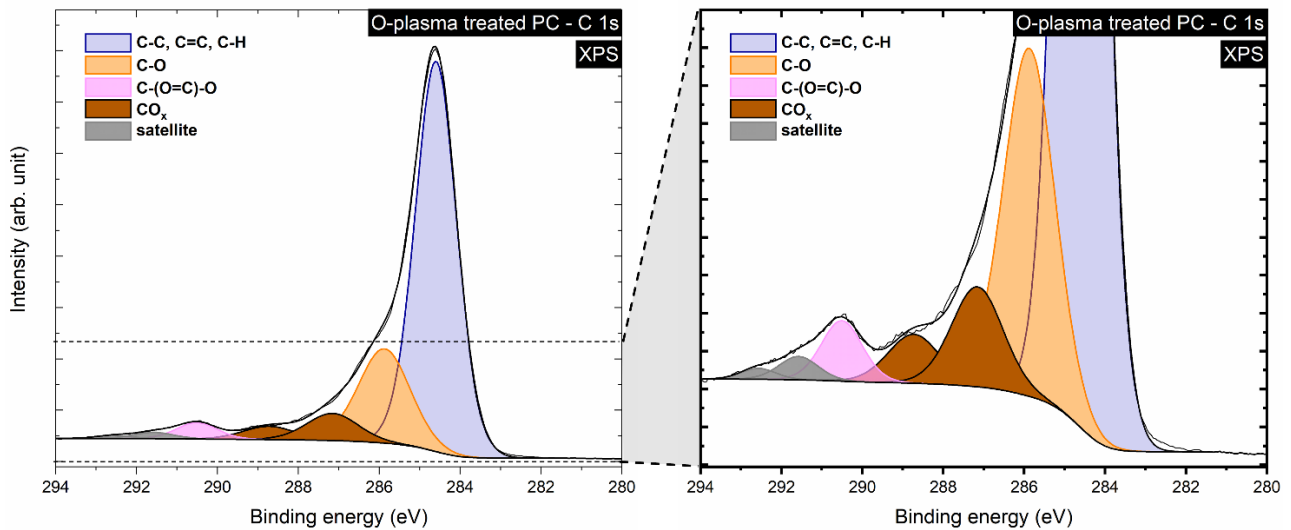

**Figure S1.** XPS C 1s spectra of O-plasma-treated PC surface ($t_{treatment}$ = ~ 1 s, $P_{pulsed\,DC}$ = 50 W, $f$ = 250 kHz, $t_{OFF}$ = 1616 ns, $p_{O2}$ = 0.72 Pa) shown as (left) whole spectra and (right) magnification of the low-intensity C 1s components. The different groups are indicated by the color code referenced in the legend.

## 7.2. XPS C 1s and O 1s spectra of the PC | TiAlO2 interface

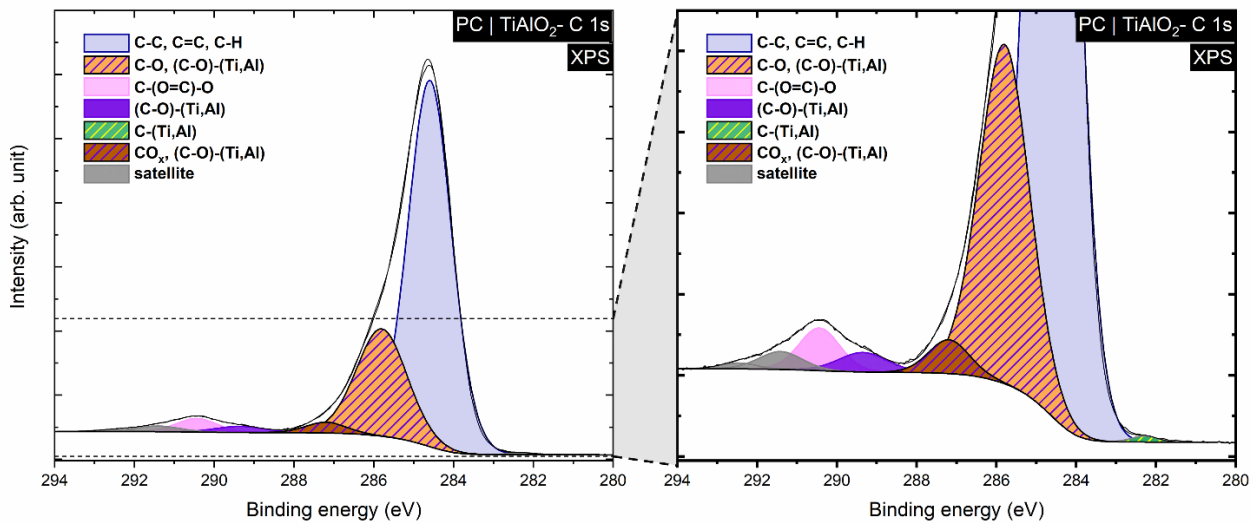

**Figure S2.** XPS C 1s spectra of the PC | TiAlO$_2$ interface are shown (left) as whole spectra and (right) as a magnification of the low-intensity C 1s components. The nominal film thickness of TiAlO$_2$ is ~ 1.0 nm. The different groups are indicated by the color code referenced in the legend.

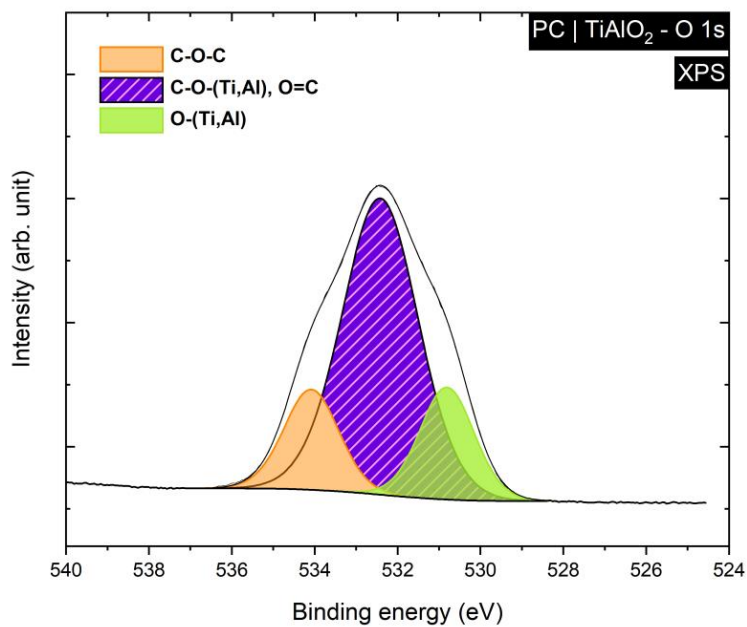

**Figure S3.** XPS O 1s spectra of the PC | TiAlO$_2$ interface after a deposition time of 0.5 s (nominal film thickness ~ 0.2 nm). The different groups are indicated by the color code referenced in the legend.

### 7.3. Calculation of the *indicator for adhesion*

**Table S1.** Average absolute ICOHP (including the statistical uncertainty as standard deviation) determined from data points in **Figure 7** [32], the *C 1s interfacial component fraction* determined from XPS spectra shown in **Figure 6**, and the resulting *indicator for adhesion* for different interfacial groups (including the statistical uncertainty as standard deviation):

    Average absolute ICOHP (column 2)
×    *C 1s interfacial component fraction* (column 1)
=    *Indicator for adhesion* (column 3)

|  | (C-O)-metal | | | C-metal | | |
|---|---|---|---|---|---|---|
|  | C1s component fraction (%) | Average absolute ICOHP (eV) | Indicator for adhesion (% × eV) | C1s component fraction (%) | Average absolute ICOHP (eV) | Indicator for adhesion (% × eV) |
| PC \| $Al_2O_3$ | 22.92 | 10.38 ± 1.38 | 237.90 ± 31.63 | 0.19 | 2.64 ± 1.23 | 0.50 ± 0.23 |
| PC \| $TiO_2$ | 5.66 | 13.05 ± 2.32 | 73.86 ± 13.13 | 0.31 | 2.31 ± 1.06 | 0.72 ± 0.32 |
| PC \| $TiAlO_2$ | 8.39 | 11.71 ± 1.33 | 98.25 ± 11.16 | 0.21 | 2.39 ± 0.08 | 0.50 ± 0.02 |
|  | (1a) | (2a) | (3a) | (1b) | (2b) | (3b) |